\documentstyle[preprint,aps,epsf]{revtex}
\begin{document}
\title{Structure functions and energy dissipation 
dependence on Reynolds number}

\author{G.~Boffetta$^{1}$ and  G.P.~Romano$^{2}$}
\address{$^{1}$ Dipartimento di Fisica Generale and INFM,
Universit\`a di Torino, \\
Via Pietro Giuria 1, I-10125 Torino, Italy}
\address{$^2$ Dipartimento di Meccanica ed Aeronautica,
Universit\`a di  Roma "la Sapienza", \\
Via Eudossiana 18, I-00184 Roma, Italy}
\date{\today}

\maketitle

\begin{abstract}
The dependence of the statistics of energy dissipation on the Reynolds
number is investigated in an experimental jet flow.
In a range of about one decade of $Re_{\lambda}$ (from about 200 to 2000)
the adimensional mean energy dissipation is found to be independent on 
$Re_{\lambda}$, while the higher moments of dissipation show a power-law 
dependence. The scaling exponents are found to be consistent with a simple
prediction based on the multifractal model for inertial range structure
functions. 
This is an experimental confirmation of the connection between inertial
range quantities and dissipation statistics predicted by the 
multifractal approach.
\end{abstract}


\section{Introduction}
\label{sec:1}

There is considerable experimental \cite{AGHA84} and numerical 
\cite{VM91} evidence that the
longitudinal velocity difference structure functions
\begin{equation}
S_{q}(\ell) \equiv \langle \delta u(\ell)^q \rangle = 
C_{q} \langle \varepsilon \rangle^{q/3} \ell^{q/3}
\left({\ell \over L} \right)^{\zeta_q-q/3}
\label{eq:1}
\end{equation}
are affected by intermittency corrections in the
scaling exponents $\zeta_q$ which deviate from the Kolmogorov 
self-similar prediction $\zeta_q=q/3$ \cite{MY75,F95} ($C_{q}$ is a 
constant possibly depending on the Reynolds number, $\varepsilon$ is 
the turbulent kinetic energy dissipation, $\ell$ is a generic length scale 
and $L$ is the integral length scale).
The estimate of $\zeta_q$ is now available for a wide range of
Reynolds numbers and different flow configurations. 
The analysis of experimental data show that structure functions
display the scaling behavior (\ref{eq:1}) for sufficiently high Reynolds number, 
$Re$, and that in such conditions the longitudinal exponents 
$\zeta_p$ become independent on $Re$ 
\cite{AZX00,APZ00}. There is also experimental support for
the scaling exponents to be $Re$-independent also at small $Re$
(when the scaling behavior in (\ref{eq:1}) is even not observable)
if one make use of the so-called extended self similarity analysis 
(ESS) \cite{ABBB96}.
However, the precise relation between this empirical result and the classical 
scaling behavior (\ref{eq:1}), expected from dimensional arguments,
is not well understood. We will not further discuss this point here.

The basic fundamental property of fully developed turbulence
is that the average energy dissipation $\langle \varepsilon \rangle$ 
entering in (\ref{eq:1}) is asymptotically $Re$-independent,
when adimensionalized with large scale variables \cite{MY75,S84,S98}.
On the other hand, it is well known that the statistics of the
local energy dissipation, which assuming isotropy can be defined in terms
of its $1D$ surrogate
\begin{equation}
\varepsilon(x) = 15 \nu \left( {\partial u \over \partial x} \right)^2 \, ,
\label{eq:2}
\end{equation}
strongly depends on $Re$, becoming more and more intermittent with
increasing $Re$ \cite{MS87,MS91}. This effect is also reflected in the 
tails of the pdf of velocity increments at very small separations
\cite{BBPVV91,KSS92,TZBMW96}.
A statistical description of the behavior of the local energy dissipation 
(\ref{eq:2}) can be obtained in terms of quantities defined in the 
inertial range, such as (\ref{eq:1}), within the multifractal approach.
This approach, originally introduced as a phenomenological model for 
the inertial range statistics \cite{PF85,PV87}, has been 
extended to the prediction of dissipative scale statistics 
\cite{F95,FV91}.
In particular, this approach predicts a scaling behavior of the moments 
of (\ref{eq:2}) with $Re$ \cite{N90}, which has been recently measured in
a simplified model of turbulence \cite{BCR00}.

In the present paper we investigate the Reynolds dependence of the 
statistics of (\ref{eq:2}) in a experimental water jet.
We have examined a series of data obtained from LDA measurements
which cover about one decade of Reynolds numbers. 
The quality of the statistics allows us to compute high order 
structure functions with reasonable accuracy and
to partially resolve the dissipative scales.
Our main result is that the dependence of the statistics of 
dissipation on the Reynolds number
is found to be consistent with the multifractal 
prediction obtained by assuming a fluctuating dissipative scale.

The remaining of the paper is organized as follows.
In Section~\ref{sec:2} we introduce the theoretical 
models for the statistics of dissipation. In Section~\ref{sec:3}, 
the experimental set-up and the data analysis procedure.
Section~\ref{sec:4} is devoted to the presentation of the results, whereas
concluding remarks are given in Section~\ref{sec:5}.

\section{Multifractal description of energy dissipation}
\label{sec:2}

A phenomenological description of intermittency is given by the
multifractal model \cite{PF85}. This model introduces a
continuous set of scaling exponents $h$ which locally 
relates the velocity fluctuations at scale $\ell$ entering 
in (\ref{eq:1}) with a large scale velocity fluctuation $u'$:
\begin{equation}
\delta u(\ell) \sim u' \left({\ell \over L} \right)^{h} \, .
\label{eq:3}
\end{equation}
The local exponent $h$ is realized with a probability 
which scales with $(\ell/L)^{Z(h)}$ where $Z(h)$ is the 
codimension of the fractal set on which the $h$-scaling
holds. The scaling exponent of structure functions (\ref{eq:1})
are obtained by a steepest descent argument over $h$:
\begin{equation}
\zeta_q = \inf_{h} \left[qh + Z(h) \right] \, .
\label{eq:4}
\end{equation}

The scaling region of (\ref{eq:3}) is bounded from below
by the Kolmogorov dissipative scale $\eta$ at which dissipation
starts to dominate, i.e. at which the local Reynolds number is of order
$1$:
\begin{equation}
{\eta \delta u(\eta) \over \nu} \simeq 1 \, .
\label{eq:5}
\end{equation}
From (\ref{eq:3}) and (\ref{eq:5}) one obtains that in the
multifractal description of intermittency the dissipative scale
is a fluctuating quantity, i.e. depends on the local 
scaling exponent $h$ according to
\begin{equation}
\eta \sim L \left({u' L \over \nu} \right)^{-{1 \over 1+h}} \sim
L Re_{\lambda}^{-{2 \over 1+h}}
\label{eq:6}
\end{equation}
where $Re_{\lambda}=u' \lambda/\nu$ is the Reynolds number based 
on the Taylor microscale $\lambda=\sqrt{15 \nu u'^2/\langle \varepsilon 
\rangle}$.

Below the dissipative scale the flow can be assumed smooth and
one can replace the derivative in (\ref{eq:2}) with
\begin{equation}
\varepsilon = 15 \nu \left( {\delta u(\eta) \over \eta} \right)^2 \, .
\label{eq:7}
\end{equation}
Assuming that the multifractal model can be pushed down to the dissipative
scale, one can evaluate the statistics of (\ref{eq:2}) by inserting
(\ref{eq:3}) and (\ref{eq:6}) into (\ref{eq:7}). One ends with the
expression \cite{N90,F95,BCR00}
\begin{equation}
\langle \varepsilon^p \rangle \sim \langle \varepsilon \rangle^p
\int d\mu(h) Re_{\lambda}^{-2 [3 p h - p + Z(h)]/(1+h)} 
\sim \langle \varepsilon \rangle^p Re_{\lambda}^{- 2 \theta_p}
\label{eq:8}
\end{equation}
where again the integral has been evaluated by a steepest descent 
argument as
\begin{equation}
\theta_p = \inf_{h} \left[{3 p h - p + Z(h) \over 1 + h} \right]
\label{eq:9}
\end{equation}

The standard inequality in the multifractal model (following
from the exact result $\zeta_3=1$), $Z(h) \ge 1-3h$ \cite{F95},
implies for (\ref{eq:9}) $\theta_{1}=0$ which is nothing but the
request of finite non-vanishing dissipation in the limit $Re \to \infty$.
For $p > 1$, $\theta_p < 0$, i.e. the tail of the distribution
of $\varepsilon$ becomes wider with the Reynolds number.

Let us remark that the prediction (\ref{eq:9}) is based on the
assumption of a fluctuating dissipative scale $\eta$ according to 
(\ref{eq:6}).
If one assume, on the contrary, that dissipation enters in (\ref{eq:7}) 
as an average quantity one ends up with a different 
predictions for the exponents $\theta_p$ \cite{MY75,EG92}. 
Numerical simulations with a simplified
model of turbulent cascade have shown that the exponents (\ref{eq:6})
are indeed observed and the alternative prediction is ruled out
\cite{BCR00}. In the next Section, we will see that also our experimental
data are in agreement with prediction (\ref{eq:9}).

In the following we will consider data analysis of one component of
the velocity in water jets at different Reynolds numbers. 
Because of the discretization of the
acquisition, we are forced to replace spatial derivatives 
with velocity differences at small scales.
The key quantity for our discussion is a generalization
of (\ref{eq:2}) over a finite scale $\ell$:
\begin{equation}
E(\ell) \equiv 15 \nu \left( {\delta u(\ell) \over \ell} \right)^2 \, .
\label{eq:10}
\end{equation}
$E(\ell)$ is a convenient definition of surrogate energy dissipation 
if the scale $\ell$ is sufficiently small.
The average dissipative scale dependence on Reynolds number can be
obtained from (\ref{eq:7}) and (\ref{eq:1}) as 
\begin{equation}
\bar{\eta} \simeq L Re_{\lambda}^{- {2 \over 2- \zeta_2}} \, .
\label{eq:11}
\end{equation}
In the analysis of experimental data it is convenient to normalize
separations with $\bar{\eta}$ 
and we will consider (\ref{eq:10}) at fixed $\ell^{*}=\ell/\bar{\eta}$.

In the limit of very small separations (i.e. $\ell^* \simeq 1$)
(\ref{eq:10}) recovers the finite difference representation of
the 1D-energy dissipation (\ref{eq:7}) and thus, from (\ref{eq:8})
\begin{equation}
{\langle E(\ell)^p \rangle \over {\langle E(\ell) \rangle}^p} 
= {\langle \varepsilon^p \rangle \over {\langle \varepsilon \rangle}^p}
\simeq Re_{\lambda}^{-2 \theta_p}
\label{eq:12}
\end{equation}
with the exponents $\theta_p$ given by (\ref{eq:9}).

In the case of separations in the inertial range, 
from (\ref{eq:1}) we have
$\langle E(\ell)^p \rangle/\langle E(\ell) \rangle^p \simeq 
\ell^{\zeta_{2p}-p \zeta_{2}}$. 
Because $\ell=\ell^{*} \bar{\eta}$, using (\ref{eq:11}) we end with
the prediction
\begin{equation}
{\langle E(\ell)^p \rangle \over \langle E(\ell) \rangle^p} \simeq
Re_{\lambda}^{-{2 \zeta_{2p} - 2 p \zeta_{2} \over 2 - \zeta_{2}}}
\label{eq:13}
\end{equation}

Let us remark that the exponents in (\ref{eq:12}) and (\ref{eq:13})
are expected to be not very different (they are both zero for 
non-intermittent turbulence) and thus a discrimination between the
two prediction require good accuracy.

\section{Experimental set-up and data analysis}
\label{sec:3}

\begin{figure}[ht]
\epsfbox{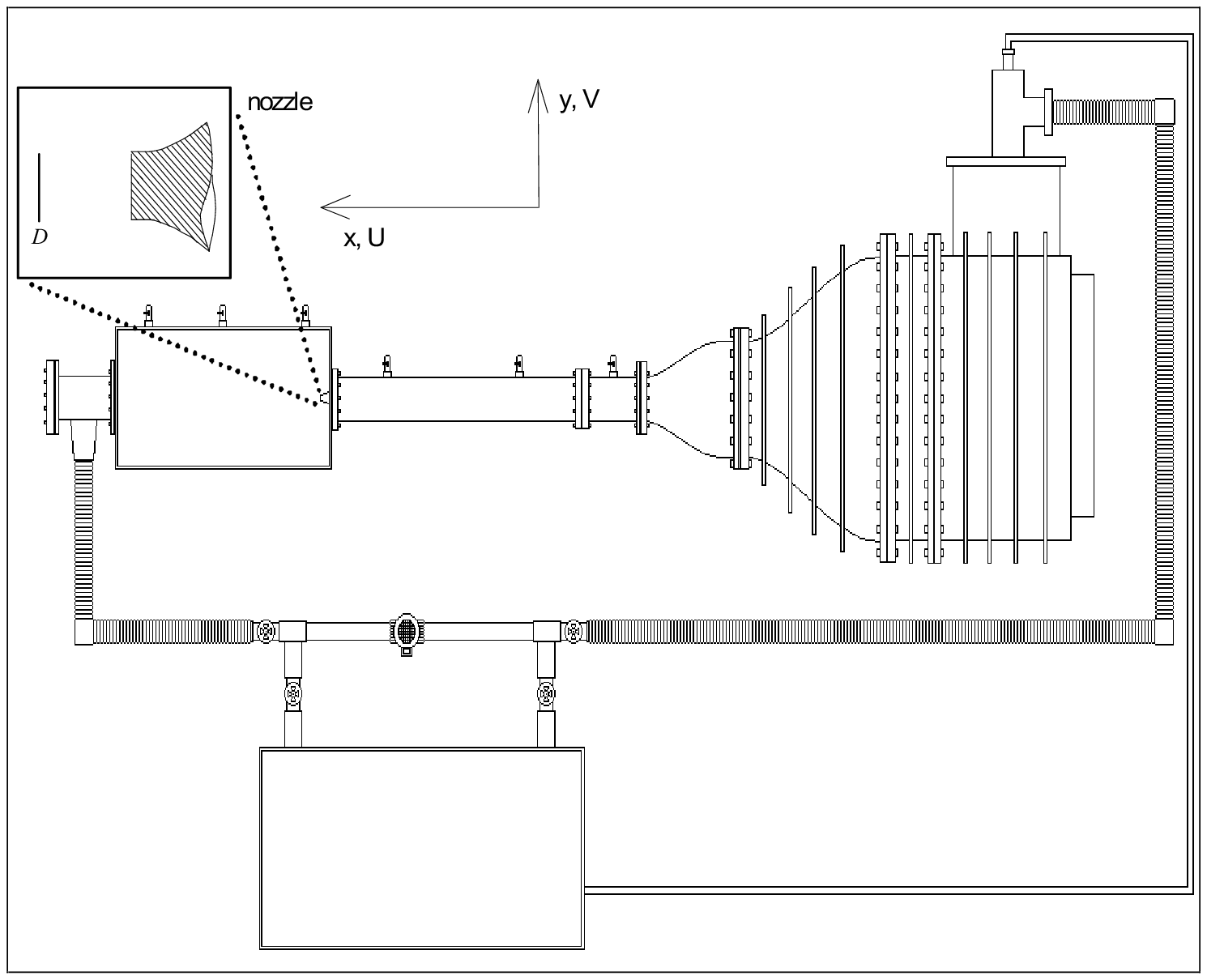}
\caption{
Sketch of the experimental setup.
}
\label{fig1}
\end{figure}

The experimental setup consists of a water jet in a closed circuit 
facility as shown in Figure~\ref{fig1}. A centrifugal pump moves water from a
primary tank into a settling chamber which is equipped with a valve to
damp oscillations due to the pump.  A series of contractions leads to 
a 1.5 m long pipe, with an inner
diameter of 14 cm which is followed by the final contraction (1:50 in
area) to the jet. The jet (diameter $d = 20$ mm) exits into a
large water-filled tank (height $30d$, width $30d$, length $60d$) from
which the water returns to the primary tank. The pipe, the contraction and
large tank are made of perspex to allow optical access to the
flow.  At the jet exit, the flow is axisymmetric and has no swirl;
preliminary measurements also confirmed that it is unaffected by any
external forcing due to the pump. The jet has a top-hat velocity
profile at the nozzle exit with a boundary layer shape parameter (defined 
as the ratio between the displacement and momentum thickness at the 
outlet) equal to $3.29$ and a turbulent intensity equal to about 
$0.021$ on the jet axis.

Velocity measurements are performed by means of a forward-scatter
Laser Doppler Anemometer (LDA) equipped with two
Bragg cells.  The fringe spacing is $3.416~\mu m$ and the measurement
volume size is about $0.1~mm$, $0.1~mm$, $0.8~mm$ along the $x$, $y$ and $z$
axes respectively.  The LDA data are randomly distributed in time; 
therefore, they are resampled by using a linear
interpolation to obtain evenly spaced samples and also to provide unbiased
statistics {\cite{RA01}}.  
The value of $Re_{\lambda}$ is sufficiently large to expect
the LDA noise not to affect the behavior of structure functions in the
IR \cite{AZR97}.  The LDA data on longitudinal and transverse structure 
functions and on scaling exponents were compared to data obtained by Hot 
Wire Anemometry in a similar jet; the agreement was good up to 
$8^{th}$ order \cite{RA01}.

Measurements were made at $x/d \simeq 40$ ($x$ is measured
from the nozzle exit plane), where the flow field may be considered to
be approximately self-preserving and isotropic \cite{AZR97}. The jet 
exit velocity $U_0$ was
selected so that the exit Reynolds number $Re_{d} \equiv U_0 d/\nu$
changes from about $2\times 10^{4}$ to $2 \times 10^{5}$.
As a consequence, $Re_{\lambda}$ changes from about $200$ to 
almost $2000$ at the measurement location. 
The number of collected samples is about $10^{6}$ in each run.
As usually done in practice, we resort to Taylor hypothesis to 
transform time differences into space differences by means of the
local mean velocity $U$.
The number of independent samples, given by $U T_{S}/2L$ 
(where $T_{S}$ is the total record duration), is about
$10^{4}$. Probability density functions of the
longitudinal velocity increments have been calculated at different values 
of $\ell^{*}$.  The distributions indicate that
the number of samples is adequate for achieving a closure of the
integrand associated with the structure functions $S_{q}(\ell)$
at least up to $q = 8$.
In Table~\ref{table1} we summarize some parameters of the experiments.

\begin{figure}[ht]
\epsfbox{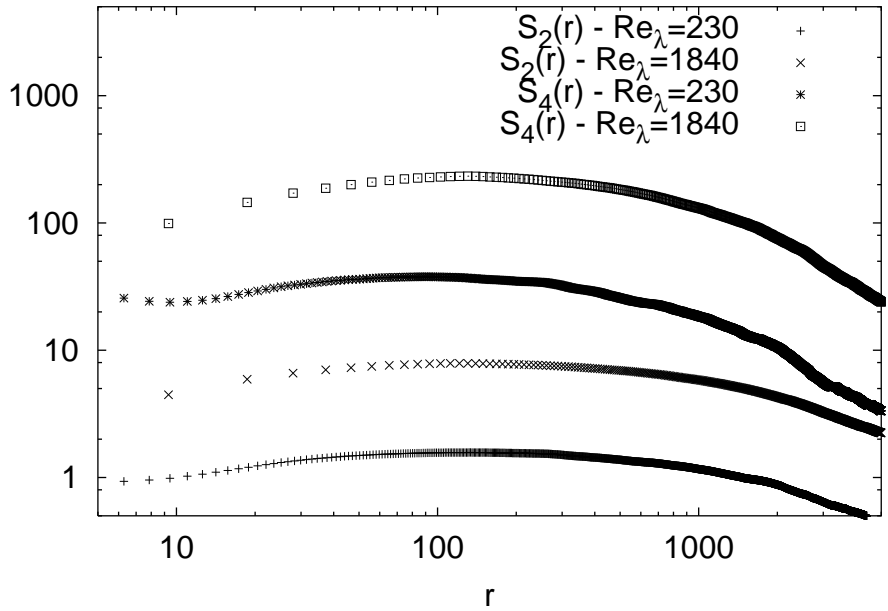}
\caption{
Structure function of order $2$ and $4$ for different values of
$Re_{\lambda}$ compensated with the scaling behavior (\ref{eq:1}).
The x-axis is adimensionalized with the Kolmogorov scale.
}
\label{fig2}
\end{figure}

Because the acquisition rate is constant for the different runs 
($\delta t=5 \times 10^{-4}~s$) the smallest resolved scale
$\delta x=U \delta t$ varies with $Re_{\lambda}$. 
By increasing $Re_{\lambda}$ we have both an increasing of 
$\delta x$ (as $Re_{\lambda}^2$) and a decreasing of the dissipative
scale $\bar{\eta}$ according to (\ref{eq:11}). 
Despite these limitations, we are confident that are able to
resolve, at least partially, the dissipative range in all the
reported runs.
In the extreme case $Re_{\lambda}=1840$ we are able to resolve 
down to $\ell^{*}\simeq 10$ which is just at the border of the 
dissipative range \cite{AP00}. In Figure~\ref{fig2}, the second and 
fourth-order structure functions obtained for $Re_{\lambda} = 230$ and 
for $Re_{\lambda}=1840$ compensated with the scaling behavior 
(\ref{eq:1}) are plotted (Kolmogorov length and velocity scales 
are used to adimensionalise the horizontal and vertical axes).
The large scale asymptotic values are also
functions of the Reynolds number and, for the second-order structure 
functions, increase almost linearly as $Re_{\lambda}$; this result 
is in agreement with that simply derived by dimensional arguments on 
{\it{rms}} and Kolmogorov velocities, i.e. that 
$u'^{2}/\langle \delta u(\eta)^2 \rangle \sim Re_{\lambda}$.
Scaling exponents can be derived from structure functions of
Figure~\ref{fig2} within 
the inertial range (approximately from $\ell \simeq 80$ to 
$\ell \simeq 500$, using a criterion based on $3\%$ difference from 
the maximum in the third-order structure function divided by $\ell$). 
The results ($\zeta_{2} = 0.708 \pm 0.020$ and 
$\zeta_{4} = 1.265 \pm 0.035$ for $Re_{\lambda}=230$) 
agree with those obtained numerically and experimentally by other 
authors \cite{F95}.

\section{Results and comparison with theoretical predictions}
\label{sec:4}

\begin{figure}[ht]
\epsfbox{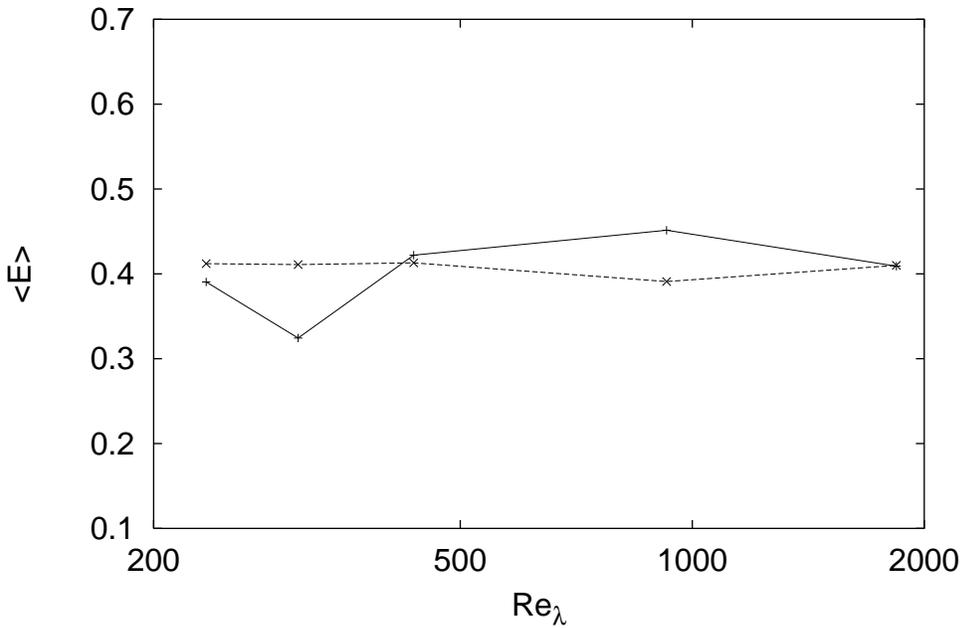}
\caption{
Average energy dissipation computed from definition (\ref{eq:10})
($+$) at $\ell^{*}=10$ and from the energy spectrum ($\times$) as function
of $Re_{\lambda}$.
}
\label{fig3}
\end{figure}

In Figure~\ref{fig3} we plot the average surrogate energy dissipation
$\langle E(\ell) \rangle$, computed from (\ref{eq:10}) and 
adimensionalized with large scale 
quantity $u'^3/L$, as a function of $Re_{\lambda}$. A
constant value of energy dissipation is expected from energy balance 
arguments \cite{MY75}. Thus, the observed independence of
$\langle E(\ell) \rangle$, computed at fixed $\ell^{*}=10$, on
$Re_{\lambda}$ is a confirmation that the dissipative
scales are resolved. 
We can give a different estimation of the dissipation on the basis
of the energy spectrum. The advantage in this case is that
we can reduce the noise by interpolation of the spectrum down to 
the dissipative scales \cite{MY75,AP00}.
The result for our data, also plotted in Figure~\ref{fig3}, is 
consistent with the
previous one thus confirming that the evaluation of the energy
dissipation statistics from the present data is reliable.

\begin{figure}[ht]
\epsfbox{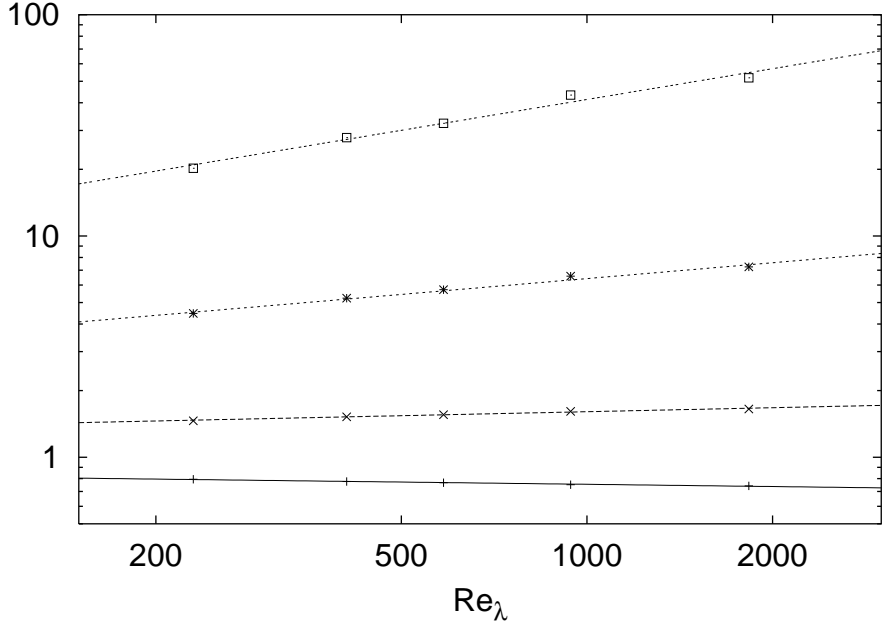}
\caption{
$Re_{\lambda}$ dependence of $\langle E(\ell)^p \rangle$ 
normalized with $\langle E(\ell) \rangle^p$
for $p=2/3$ ($+$), $p=4/3$ ($\times$), $p=2$ ($*$) and $p=8/3$ 
($\Box$) at $\ell=10 \eta$. The lines represent the best fit 
with a power law.
}
\label{fig4}
\end{figure}

In order to verify the prediction for $\theta_p$ of Section~\ref{sec:2}, the 
scaling laws (\ref{eq:12}) of the moments of $E(\ell)$ with $Re_{\lambda}$
must be computed. In Figure~\ref{fig4}, an example of this scaling,
computed at $\ell^{*}=10$, is given for different values of $p$. 
We are confident that at least up to $p=3$ the scaling 
exponents can be evaluated with an error on the fit less than $5\%$.

\begin{figure}[ht]
\epsfbox{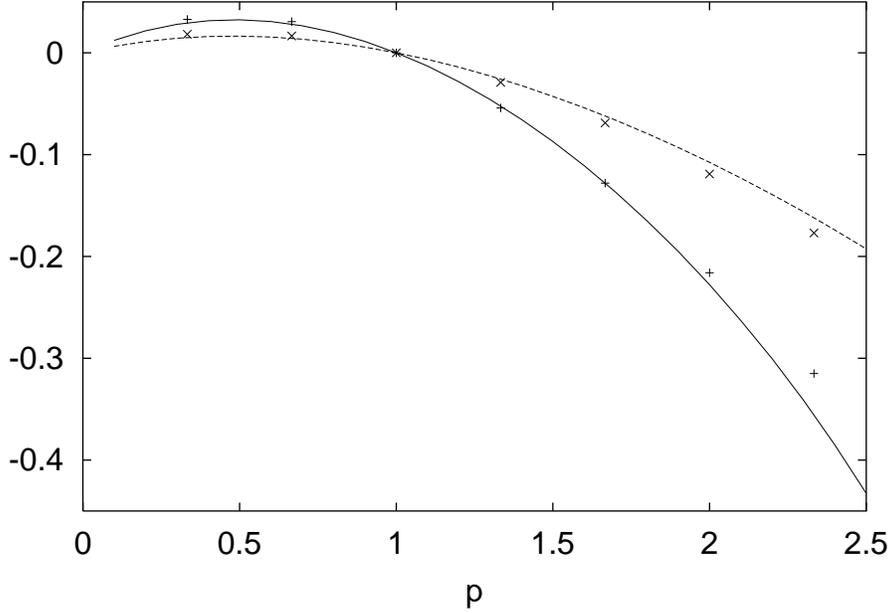}
\caption{
Exponents $\theta_p$ obtained from the data of Figure~\ref{fig4}
for $\ell=10 \eta$ ($+$) and $\ell=100 \eta$ ($\times$).
The continuous line is the intermittent prediction (\ref{eq:9}),
the dashed line is the prediction (\ref{eq:13}). 
}
\label{fig5}
\end{figure}

The resulting exponents $\theta_p$ are given in Figure~\ref{fig5}.
They have been computed at two different values of the separation 
$\ell^{*} =10, 100$. As expected, for separations moving from 
the dissipative range into the inertial range, the curve of the 
exponents $\theta(p)$ becomes flatter and approaches the prediction
(\ref{eq:13}). In Figure~\ref{fig5} we also plot the two predictions
(\ref{eq:9}) and (\ref{eq:13}). These are been obtained by computing the
structure function scaling exponents $\zeta_{q}$ (\ref{eq:1}) as discussed
above, at the largest $Re_{\lambda}$ available. 
Given the exponents $\zeta_{q}$, the codimension $Z(h)$ has
been obtained by numerical inversion of (\ref{eq:4}). Finally, 
(\ref{eq:9}) is used to predict the values of $\theta_p$.

The agreement of our data with the two predictions (\ref{eq:12})
and (\ref{eq:13}) is remarkable. 
The value of $\theta_p$ obtained from $E(\ell)$ at $\ell^{*} =10$
are very close to prediction (\ref{eq:9}-\ref{eq:12}). Some deviations are
observed, starting from moment $p \simeq 2.5$. These are probably due
to the relatively large value of $\ell^{*} =10$, which is at the 
border of the dissipative range and to the statistics. 
Indeed, according to (\ref{eq:10}), the computation of $\theta_{3}$
corresponds roughly to the computation of the $6$-th order structure 
function.

Let us remark again that the difference between predictions 
(\ref{eq:9}) and (\ref{eq:13}) is based on the assumption
of a dissipative scale fluctuating with the local energy dissipation.
Our experimental data demonstrate that this is indeed the case.

\section{Remarks and conclusions}
\label{sec:5}

In this paper, the scaling of velocity increments at very small scales
in a turbulent flow is investigated with special focus on Reynolds number 
dependence.
Within the multifractal framework it is possible to derive a relation 
between the scaling of the energy dissipation statistics with the 
Reynolds number and the scaling exponents of the velocity structure 
functions at fixed Reynolds number. This relation is found to fit very well 
experimental data taken into a water jet over about a decade of 
$Re_{\lambda}$. The one-dimensional surrogate of energy dissipation 
is estimated from velocity differences at very small scale, at
the border of the dissipative range.
It is shown that our statistics changes from prediction (\ref{eq:12})
to prediction (\ref{eq:13}) as the separation $\ell$ moves into the
inertial range. The difference between the two predictions is the
demonstration of the fluctuation of the dissipative scale.

A Reynolds number dependence of the moments of the velocity increments 
also in the limit of very high values ($Re_{\lambda} > 1000$) has 
been recently reported by different authors \cite{APZ00}. 
Due to the fact that the 
Reynolds number is expected to affect mainly the dissipative scales, 
the previous findings confirm experimentally the connection between small 
and inertial range scalings. This could support ESS (for example in 
respect to the enhanced range of scales involved), but simultaneously 
invalidate it (in respect of the independence on the Reynolds number).
If these results are dependent on the still finite value of the Reynolds 
number or on the anisotropy of the flow field is left as a subject for
future investigations.


\newpage
\begin{table}
\begin{tabular}{|c||c|c|c|c|c|}
Outlet mean velocity &&&&& \\
$U_{0}$ (cm/s) & 102 & 196 & 304 & 487 & 1060 \\
\hline
$Re_{d}$ & 20400 & 39200 & 60800 & 97400 & 212000 \\
\hline
Local mean velocity &&&&& \\
$U$ (cm/s) & 15.3 & 30.4 & 41.4 & 72.1 & 167 \\
\hline
Local {\it rms} velocity &&&&& \\
$u'$ (cm/s) & 4.4 & 9.2 & 12.3 & 23.8 & 50.1 \\
\hline
$Re_{\lambda}$ & 230 & 308 & 435 & 926 & 1840 \\
\hline
Local integral scale &&&&& \\
$L$ (cm) & 5.5 & 6.3 & 9.4 & 22.0 & 41.1 \\
\hline
Local Kolmogorov scale &&&&& \\
$\eta$ (cm) & 0.0181 & 0.0097 & 0.0086 & 0.0066 & 0.0045 \\
\end{tabular}
\caption{
Parameters of the experiments on the jet at $x/d = 40$ 
(except for the outlet mean velocity and the exit Reynolds number 
computed at $x/d = 0$).}
\label{table1}
\end{table}

\end{document}